\documentclass[reprint,aps,prb,floatfix,twocolumn,superscriptaddress,showpacs,amsmath,amssymb,showpacs,]{revtex4-1} 

\usepackage{graphicx}
\usepackage{epstopdf}
\usepackage{epsfig}
\usepackage{epsf}
\usepackage{url}
\usepackage[USenglish]{babel}
\usepackage{hyperref}
\def\bcen{\begin{center}}
\def\ecen{\end{center}}
\allowdisplaybreaks
\renewcommand\[{\begin{equation}}
\renewcommand\]{\end{equation}}
\usepackage{verbatim}
\usepackage{xcolor}
\usepackage{amsmath, nccmath}
\usepackage{bm}
\usepackage{bbm}
\usepackage{lipsum}
\usepackage{stmaryrd}
\usepackage{wrapfig}
\usepackage{soul}
\begin{document}
\title{Photo-induced charge dynamics in 1$T$-TaS$_2$}
\author{Francesco Petocchi}
\affiliation{Department of Physics, University of Fribourg, 1700 Fribourg, Switzerland}
\affiliation{Department of Quantum Matter Physics, University of Geneva, 1211 Geneva 4, Switzerland}
\author{Jiyu Chen}
\affiliation{Department of Physics, University of Fribourg, 1700 Fribourg, Switzerland}
\author{Jiajun Li}
\affiliation{Department of Physics, University of Fribourg, 1700 Fribourg, Switzerland}
\affiliation{Paul Scherrer Institute, Condensed Matter Theory, 5352 PSI Villigen, Switzerland}
\author{Martin Eckstein}
\affiliation{Institute of Theoretical Physics, University of Hamburg, 20355 Hamburg, Germany}
\author{Philipp Werner}
\affiliation{Department of Physics, University of Fribourg, 1700 Fribourg, Switzerland}

\begin{abstract}
Recent theoretical studies showed that the electronic structure of 1$T$-TaS$_2$ in the low-temperature commensurate charge density wave phase exhibits a nontrivial interplay between band-insulating and Mott insulating behavior. This has important implications for the interpretation of photo-doping experiments. Here we use nonequilibrium dynamical mean-field theory simulations of a realistic multi-layer structure to clarify the charge carrier dynamics induced by a laser pulse. The solution is propagated up to the picosecond timescale by employing a memory-truncation scheme. While long-lived doublons and holons only exist in the surface state of a specific structure, the disturbance of bonding states in the bilayers which make up the bulk of the system explain the almost instantaneous appearance of in-gap states. Our simulations consistently explain the coexistence of a doublon feature with a prominent ``background" signal in previous time-resolved photoemission experiments, and they suggest strategies for the selective population of the ingap and doublon states by exploiting the sensitivity to the pump polarization and pump frequency.  
\end{abstract}
\date{\today}
\maketitle
%
%
%
%
\section{INTRODUCTION}
The layered transition metal dichalcogenide $1T$-TaS$_2$ exhibits a complex equilibrium phase diagram, which includes incommensurate, nearly commensurate, and commensurate charge density wave (CDW) orders at ambient pressure. In the low-temperature commensurate CDW (CCDW) state, which is formed below 180 K, the material becomes insulating and exhibits a periodic lattice distortion.\cite{Wilson1975} This distortion leads to the formation of so-called star-of-David (SOD) clusters consisting of 13 Ta atoms. In a monolayer of  $1T$-TaS$_2$, these clusters form a triangular lattice and the bandstructure exhibits a half-filled narrow band near the Fermi level.\cite{Pasquier2021} The Coulomb interaction then leads to a splitting of this band into upper and lower Hubbard bands, which has been confirmed experimentally.\cite{Lin2020} For many years, $1T$-TaS$_2$ in the CCDW state has thus been considered to be a (polaronic) Mott insulator.\cite{Fazekas1979} This view has however been challenged by recent experimental\cite{Cho2016,Butler2020,Lee2021,Wu2022,Gerasimenko2019} and theoretical\cite{Darancet2014,Ritschel2015,Ritschel2018,Lee2019,Shin2021,Lee2021,Petocchi2022} works, which highlight the importance of bilayer substructures. According to density functional theory (DFT) calculations, the lowest energy structure exhibits a specific stacking of bilayers,\cite{Lee2019} which leads to a different interpretation of the bulk insulating properties in terms of a bonding/antibonding gap. For a termination of the system within the bilayers, one would get a metallic surface state within DFT, while correlations turn this surface state into a Mott insulator. If the termination occurs between bilayers, the entire structure is band insulating.\cite{Petocchi2022} Indeed, recent scanning tunneling microscopy\cite{Butler2020,Lee2021}  and photoemission experiments\cite{dong2022} can be consistently explained by assuming a coexistence of domains with band insulating and Mott insulating behavior near the surface. 

This evolution in the theoretical understanding of the electronic structure of  $1T$-TaS$_2$ has profound implications for the interpretation of nonequilibrium experiments.\cite{Perfetti2006,Ligges2018,dong2022} In particular, an analysis based on a simple Mott insulator picture cannot adequately describe photo-induced states. $1T$-TaS$_2$ has been extensively studied with pump-probe techniques, and it has attracted a lot of attention after the discovery of light-induced hidden phases.\cite{Stojchevska14,Cho2016}

In this study, we will consider photo-excitation protocols which do not change the structure of the CCDW phase, and we mainly focus on electronic processes which are fast compared to the 400 fs oscillation period\cite{Perfetti2006} of the SOD breathing mode. Our aim is to clarify how the interplay between Mott insulating surface states and band insulating bilayers affects the photo-induced charge dynamics, using a dynamical mean-field theory (DMFT)\cite{Georges1996} based modeling of a realistic multi-band structure. This analysis provides a consistent interpretation of the experimental results reported in Refs.~\onlinecite{Ligges2018} and \onlinecite{dong2022}, which showed the appearance of two qualitatively different features after photo-excitation across the gap: (i) a relatively narrow peak at an energy near 0.2 eV, which was associated with the creation of doublons (doubly occupied sites), and (ii) an unexplained broad spectral feature which fills in the gap region and extends beyond the doublon peak. In particular, we will reveal that the doublon feature originates entirely from the Mott insulating surface state, while the in-gap states are the result of a photo-induced ``melting" of the bilayers. 

With the use of a memory truncation scheme in the solution of the DMFT equations, the computational effort scales linearly in the maximum simulation time, and the solution can be propagated up to the picosecond timescale. This allows us to demonstrate that the photo-induced doublons, which are pinned to the Mott insulating surface state, decay on a timescale which is several orders of magnitude longer than the life-time of the excitations created within the bilayers.

The paper is structured as follows. In Sec.~\ref{sec:method} we describe the dynamical mean field theory based method used for the semi-realistic description of  photo-excited $1T$-TaS$_2$ multi-layer structures. Section~\ref{sec:results} presents the results for the charge carrier dynamics and the nonequilibrium spectra, as well as an analysis of the sensitivity to the polarization and frequency of the pump pulse. In Sec.~\ref{sec:conclusions} we connect the theoretical results to the existing photoemission data and present the conclusions.    
%
%
%
%
\section{METHOD} \label{sec:method}
\subsection{Inhomogeneous cluster DMFT formalism} 
\begin{figure}
\begin{center}
\includegraphics[width=0.4\textwidth]{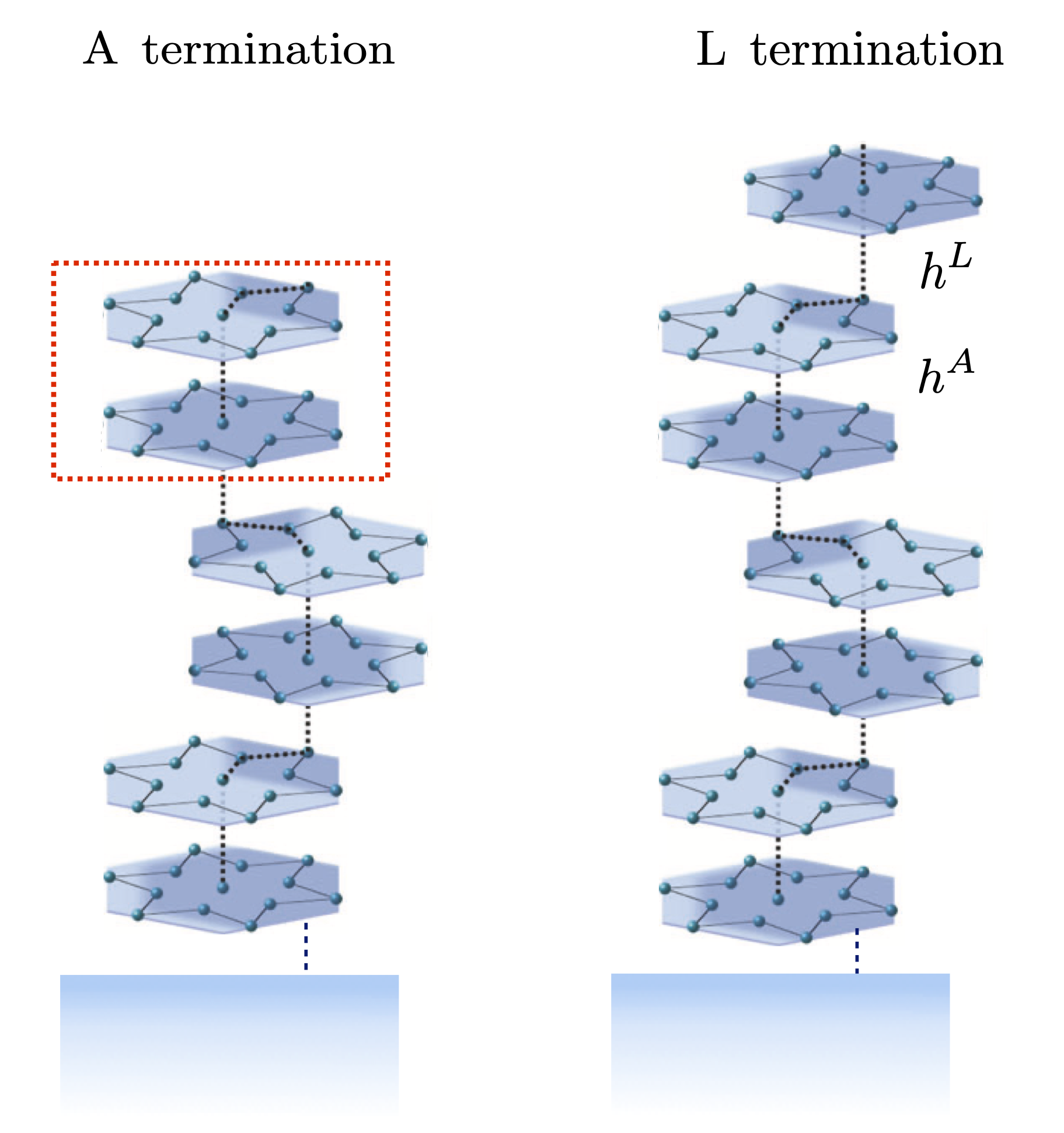}
\caption{ Sketch of the stacking arrangements of the SOD clusters located on different planes with the two possible terminations. The figure shows only a fraction of the 16 layers considered in this work. The shaded region at the bottom represents the semi-infinite bulk. A two-site DMFT cluster associated with a bilayer is indicated by the red box.\label{Figure: sketch}}
\end{center}
\end{figure}

The model Hamiltonian considered in this work is based on a realistic multi-layer description of 1$T$-TaS$_2$ in the CCDW phase, which we used in a previous equilibrium study of the electronic structure.\cite{Petocchi2022} The different layers are arranged in the vertical direction so as to reproduce a particular stacking of bilayers that has been identified as the structural ground state,\cite{Lee2019} see Fig.~\ref{Figure: sketch}. In this so-called AL stacking, the bilayers feature vertically aligned SOD clusters, while a shift along the primitive vector $\mathbf{a}$ occurs between two neighboring bilayers. The resulting stacking vector is $\mathbf{T}=-2\mathbf{a}+2\mathbf{c}$. The effective single-orbital Hamiltonian $\mathcal{H}_{\mathrm{SOD}}$ for the SOD clusters of a single layer was obtained with DFT and maximally localized Wannier orbitals, as described in Ref.~\onlinecite{Petocchi2022}, while neighboring layers are connected via two hopping parameters: $h^\text{A}=0.2$ eV for hopping processes within a bilayer and $h^\text{L}=0.045$ eV for those between bilayers. These values allow to reproduce the surface spectral functions measured with STM. Following Ref.~\onlinecite{Petocchi2022} the 
hopping part of the Hamiltonian can be written in the form
\begin{equation}
\mathcal{H}_{ab}\left(\mathbf{R}\right)=\mathcal{H}_\text{SOD}\left(\mathbf{R}\right)\delta_{ab}-h_{ab}^{A}\delta_{\mathbf{R}}-h_{ab}^{L}\left(\mathbf{R}\right),
\label{hamilt}
\end{equation}
where $\mathbf{R}$ is the unit cell vector and the index $\left\{a,b\right\}$ refers to the layer inside the unit cell. In our previous investigation we employed $GW$+EDMFT\cite{Boehnke2016,Nilsson2017,Petocchi2022} to study how the combined effect of strong correlations and strong inter-layer hybridization controls the equilibrium electronic structure. The bulk of 1$T$-TaS$_2$ was found to be a band insulator, while a Mott insulating surface state appears when the structure terminates with a broken bilayer (L termination). Despite the different nature of the two insulating states, for realistic interaction and hopping parameters, similar gaps were obtained in the bulk and near the surface, which partly explains the debate about whether or not 1$T$-TaS$_2$ should be regarded as a  Mott or band insulator.

An interesting question is if the physically different mechanisms underlying the gap formation in the equilibrium spectrum can be disentangled in the time domain, through characteristic signatures in the dynamics of photo-induced carriers. As we will show in the results section, this is indeed the case. In particular, doublons and holons produced in the Mott surface state have a longer life-time than the particle-hole excitations created within a bilayer. 

To study the interacting system in the time domain we use the nonequilibrium generalization of DMFT\cite{Aoki2014} in combination with a solver based on the strong coupling expansion (non-crossing approximation, NCA).\cite{Keiter1971,eckstein2010} Given the complexity of the system, which features several sites (layers) in a unit cell, we adopt a simplified self-consistency which is adequate for insulating systems. This approach is based on a Bethe-lattice-inspired real-space construction of the hybridization function\cite{Werner2017,Petocchi2019Hund} and has the advantage of being economical in terms of memory requirement, at the cost of losing the information on the details of the energy dispersion. 

The time-dependent hybridisation function for site $i$ can be written as
\begin{equation}
\hat{\Delta}_{i}\left(t,t'\right)=\underset{j}{\sum}\hat{h}_{ij}\left(t\right)\hat{G}_{j}^{\left[i\right]}\left(t,t'\right)\hat{h}_{ji}\left(t'\right),
\end{equation}
where $\hat{h}\left(t\right)$ is the time-dependent hopping parameter and $\hat{G}_{j}^{\left[i\right]}$ is the cavity Green's function for the lattice with the $i$th site removed. By replacing the cavity Green's function with the local Green's function -- which is an approximation in a system with finite coordination number -- one obtains a DMFT self-consistency condition relating the hybridization function directly to the local Green's functions, similar to the case of the infinitely connected Bethe lattice.\cite{Georges1996} This procedure can however not adequately capture the nonlocal correlations originating from the strong hybridization within a bilayer, which lead to the opening of the bonding/antibonding gap. For this reason, we use a two-site cluster DMFT,\cite{Lichtenstein2000} where the two sites of a cluster represent the two planes of a bilayer (Fig.~\ref{Figure: sketch}). 
In this cluster approach, the strong intra-bilayer hopping $h^\text{A}$ is exactly treated within the impurity model. 
In the non-equilibrium implementation, each bilayer is represented by a two-site impurity model with a $2\times 2$ Green's function $\hat G(t,t')$ and a $2\times 2$ hybridization function $\hat{\Delta}\left(t,t'\right)$ constructed from the DFT-derived in-plane hoppings and the weak inter-bilayer hopping $h^\text{L}$. In the following it is therefore convenient to express the real-space hoppings in terms of the bilayer index $i$ and the internal index of the layer $\alpha=\left\{1,2 \right\}$ so that $h^i_{\alpha ,\mathbf{R}}\left(t=0\right)=\mathcal{H}_\text{SOD}\left(\mathbf{R}\right)\forall \, \left\{i,\alpha \right\} $. Since $h^\text{A}$ is local and acts only within a bilayer, it will be written as $h^{i}_{12}\left(t\right)$, while the hoppings between different bilayers $h^\text{L}$, which due to the stacking vector extend also outside the unit cell, will be denoted by $h^{i,i\pm1}_{\mathbf{R}}\left(t\right)$. With this notation the in-plane hybridisation matrix for a given bilayer becomes 
\begin{equation}
\hat{\Delta}_{\Vert}^{i}=\sum_{\mathbf{R}}\left(\begin{array}{cc}
h^i_{1,\mathbf{R}}G^i_{11}h_{1,\mathbf{R}}^{*} & h^i_{1,\mathbf{R}}G_{12}h_{2,\mathbf{R}}^{i*}\\
h^i_{2,\mathbf{R}}G^i_{21}h_{1,\mathbf{R}}^{*} & h^i_{2,\mathbf{R}}G_{22}h_{2,\mathbf{R}}^{i*}
\end{array}\right), 
\end{equation}
where $G_{\alpha\beta}^{i}\left(t,t'\right)$ is the 2 $\times$ 2 contour Green's function\cite{Aoki2014} of the $i$th bilayer and the time arguments have been dropped for simplicity. With this implementation, correlation effects within each bilayer are well-captured provided that the kinetic energy from the in-plane motion of the electrons is adequately described. 

The local two-site Hamiltonian of the $i$th bilayer is given by
\begin{align}
\mathcal{H}_{i}\left(t\right)=&-\mu\left(\hat{n}_{i1}+\hat{n}_{i2}\right)+U\left(\hat{n}_{i1\uparrow}\hat{n}_{i1\downarrow}+\hat{n}_{i2\uparrow}\hat{n}_{i2\downarrow}\right) \\ \nonumber
                                           &-\sum_{\sigma}\left[h^{i}_{12}\left(t\right)\hat{c}_{i1\sigma}^{\dagger}\hat{c}_{i2\sigma}+h.c.\right], 
\end{align}
where $U$ is the local Hubbard interaction and $\mu=\frac{U}{2}$ is the chemical potential. 

The different bilayers are connected by the hopping $h^\text{L}$, which yields additional hybridization contributions related to Green's functions corresponding to adjacent layers, 
\begin{align}
\hat{\Delta}_{\bot}^{i}=\sum_{\mathbf{R}}\left(\begin{array}{cc}
h_{\mathbf{R}}^{i,i-1}G_{22}^{i-1}h_{\mathbf{R}}^{i-1,i} & 0\\
0 & h_{\mathbf{R}}^{i,i+1}G_{11}^{i+1}h_{\mathbf{R}}^{i+1,i}
\end{array}\right).
\end{align}
The total hybridization function for each bilayer is then given by the sum of the two contributions:
\begin{equation}
\hat{\Delta}^{i}=\hat{\Delta}_{\Vert}^{i}+\hat{\Delta}_{\bot}^{i}.
\label{eq_delta}
\end{equation}
Since  $h^\text{A} \gg h^\text{L}$ and the in-plane hopping is small, both the inter-bilayer and intra-bilayer hybridization contributions are small, which makes the use of the NCA meaningful. 

The solid-vacuum interface is described by setting  $h_{\mathbf{R}}^{i,i-1}=0$ for the terminating bilayer. To mimic the embedding potential from the semi-infinite bulk, we connect the bottom side of the multi-layer  setup to a non-interacting bath with the same (gapped) density of states as the last layer. We start the DMFT loop by solving separate two-site impurity problems for the different bilayers. Once all the cluster Green's functions have been computed, the hybridization functions for the different impurity problems are calculated according to Eq.~(\ref{eq_delta}) and used as input for the following DMFT iteration.

The converged solutions of the cluster impurity problems may then be used to evaluate the different contributions to the energy of the system:
\begin{align}
\epsilon_{\mathrm{intra}}^i \left(t\right) & =-\sum_{\sigma}\left\langle h_{12}^{i}\hat{c}_{i1\sigma}^{\dagger}\hat{c}_{i2\sigma}+h.c.\right\rangle \left(t\right), \label{eqn:energy} \\
u^i \left(t\right)&=\sum_{\alpha=1,2}U\left\langle \hat{n}_{i\alpha\uparrow}\hat{n}_{i\alpha\downarrow}\right\rangle \left(t\right)-\mu\left\langle \hat{n}_{\alpha}\right\rangle \left(t\right),   \\
\epsilon_{\mathrm{inter}}^i \left(t\right)  & =2\mathrm{Im}\left\{ \mathrm{Tr}\left[ \hat{\Delta}^{i} \ast \hat{G}^{i} \right] \right\}^{<} \left(t,t\right). \label{eq_ekin_inter}
\end{align}
where the trace in Eq.~(\ref{eq_ekin_inter}) is performed over the two layers of the $i$th bilayer. The kinetic energy $\epsilon_{\mathrm{intra}}\left(t\right)$ and local energy $u\left(t\right)$ of the bilayers are obtained directly from the time dependent expectation values of the intra-bilayer hopping and double occupancy. The kinetic energy $\epsilon_{\mathrm{inter}}\left(t\right)$  associated with hopping processes to other bilayers is given by the convolution between the local hybridization matrix and the local Green's function matrix. In Eq.~\eqref{eq_ekin_inter} the factor of two comes from the sum over the spin contributions. All energies are reported in eV.
\subsection{Light-matter interaction}
The coupling between the electrons and light is implemented with the Peierls substitution
\begin{equation}
\mathcal{H}_{ab}\left(\mathbf{R},t\right)=\mathcal{H}_{ab}\left(\mathbf{R}\right)e^{-\frac{ie}{\hbar}\phi_{ab}\left(\mathbf{R},t\right)}.
\end{equation}
Here, the effect of the electric field is described by the vector potential $A(t)$, whose line integral defines $\phi$. This choice is particularly convenient since the model is defined in real space and the time dependent phase factors can be included directly in the hopping parameters. The polarization $\mathbf{p}=\left\{ \mathbf{e}_{x},\mathbf{e}_{y},\mathbf{e}_{z}\right\} $ of the electric field $\mathbf{E}\left(t,z\right)=E\left(t,z\right)\mathbf{p}$ is explicitly considered and the amplitude of the pulse has a Gaussian envelope with maximum value $E_0$,
\begin{align}
E\left(t,z\right)&=E\left(t\right)f\left(z-z_{s}\right)\\ 
&=E_{0}e^{-\left(\frac{t-t_{0}}{\sigma}\right)^{2}}\sin\left(\omega_{0}t\right)f\left(z-z_{s}\right). \nonumber
\end{align}
Here, $t_0$ indicates the time corresponding to the pump maximum, $\omega_0$ the frequency of the pulse, which we express in eV hereafter, and the width at half maximum (whm) is given by $2\sigma\sqrt{\ln2}$. In all our simulations we excite only the surface region of the multi-layer structure by including a function
\begin{equation}
f\left(z-z_{s}\right)=e^{-\frac{z_{s}-z}{\gamma}}
\end{equation}
which damps the intensity of the electric field with increasing distance from the solid-vacuum interface located at $z_s$ ($z \le z_s$). 
Realistic pump pulses would excite many layers. Here, we set the characteristic length of the decay to $\gamma=0.75$. The reason for this choice is twofold: one of the goals of this study is to disentangle the local carrier recombination from their diffusion within the structure. Setting $f\left(z\right)=1$, which corresponds to a uniform excitation for all the layers, would hide the diffusion within the bulk. Secondly, exciting the layers close to the non-interacting bath mimicking the semi-inifinte bulk would result in spurious carrier recombinations. 

The Peierls phase factor is given by the integral of the vector potential 
\begin{align}
\mathbf{A}\left(t,z\right)&=-f\left(z-z_{s}\right)\int_{0}^{t}dt'E\left(t'\right)\mathbf{p}  \label{eqn:AdotP} \\
&=f\left(z-z_{s}\right)A\left(t\right)\mathbf{p} \nonumber
\end{align}
along the path corresponding to the real-space hopping:
\begin{align}
\phi_{ab}\left(\mathbf{R},t\right)=&A\left(t\right)\int_{\mathbf{r}_{a}}^{\mathbf{R}+\mathbf{r}_{b}}f\left(z-z_{s}\right)\mathbf{p}\cdot d\mathbf{r}\nonumber\\ 
=&A\left(t\right)\left\{ \mathbf{e}_{x},\mathbf{e}_{y},\mathbf{e}_{z}\right\} \cdot\left\{ \sin\theta\cos\phi,\sin\theta\sin\phi,\cos\theta\right\} \nonumber \\
&\times\int_{0}^{\ell}f\left(z-z_{s}\right)dr. 
\label{eq_phi_pol}
\end{align}
Note that for $f\left(z\right)=1$ the integral over $dr$ is equal to the absolute distance $\ell$  between the lattice sites connected by the hopping and the result $\phi_{ab}\left(\mathbf{R},t\right)=\mathbf{A}(t)\cdot\left(\mathbf{R}+\mathbf{r}_{b}-\mathbf{r}_{a}\right)$ for a homogeneous vector potential is recovered. For the exponential decay employed in this study one has
\begin{eqnarray}
\int_{0}^{\ell}e^{-\frac{z_{s}-z}{\gamma}}dr&=&\int_{0}^{\ell}e^{-\frac{z_{s}-\left(z_{a}+r\cos\theta\right)}{\gamma}}dr \label{eqn:decay}  \\
&=&\frac{\gamma}{\cos\theta}\left(e^{\frac{\ell\cos\theta}{\gamma}}-1\right)e^{-\frac{z_{s}-z_{a}}{\gamma}}, \nonumber
\end{eqnarray}
where the angle $\theta$ measures the deviation from the stacking direction. Note that the hopping within a layer corresponds to the $\cos\theta=0$ limit and the contribution from Eq.~\eqref{eqn:decay} is $\ell e^{-\frac{z_{s}-z_{a}}{\gamma}}$.  In this case the Peierls factor becomes the standard one multiplied by an exponential damping given by the distance from the surface:
\begin{equation}
\phi_{ab}\left(\mathbf{R},t\right)=\mathbf{A}(t)\cdot\left(\mathbf{R}+\mathbf{r}_{b}-\mathbf{r}_{a}\right) e^{-\frac{z_{s}-z_{a}}{\gamma}}.
\end{equation}
From Eq.~(\ref{eqn:AdotP}) one sees how the electric field polarization $\mathbf{p}$ can be used to excite hoppings in the different directions. For instance, $\mathbf{p}=\frac{1}{\sqrt{2}}\left\{ 1,1,0\right\} $ corresponds to excitations (time-dependent hoppings) within a layer, while any choice with $\mathbf{p}_z=\mathbf{e}_z\cos\theta \ne0$ will produce excitations between the layers with a field amplitude which depends on the depth inside the structure. Due to the form of the self-consistency, the model treats the in-plane motion on an approximate level. For this reason, all the results presented in this work are averaged over several different norm-conserving planar polarizations $\left\{ \mathbf{p}_x,\mathbf{p}_y\right\}$.
\subsection{Memory truncation}
The standard approach to solving the nonequilibrium DMFT equations is to discretize the time arguments with a fixed small timestep $h$ so that $t=Nh$.\cite{Aoki2014} Because this solution involves Dyson-type equations in which the self-energy plays the role of a memory kernel, a calculation which retains the memory back to the initial state is very costly in terms of memory ($\mathcal{O}(N^2)$) and computational complexity ($\mathcal{O}(N^3)$ in the NCA case). To improve the scaling and propagate the solution to longer times a memory-truncation scheme has been proposed,\cite{Schuler2018,stahl2022} in which the self-energy is stored only within a time window defined by a sufficiently large cutoff $t_{\mathrm{cut}}=N_\text{cut}h$. This allows to reduce the memory and computational costs to $\mathcal{O}(N_{\text{cut}}^2)$ and $\mathcal{O}(N\cdot N_{\text{cut}}^2)$, respectively.

In a generic nonequilibrium DMFT calculation with NCA impurity solver, two types of Dyson equations have to be solved: (i) the lattice Dyson equations, and (ii) the pseudo-particle Dyson equations of the impurity solver. In the present study, since we use the simplified Bethe-lattice-like self-consistency scheme, we can directly calculate the hybridization function $\Delta$ of the impurity action from the impurity Green's function $G_\text{imp}$ and thus circumvent the lattice Dyson equation. What remains is the NCA solution of the impurity problem which involves solutions of pseudo-particle Dyson equations. The pseudo-particle self-energies are defined at the NCA level as\cite{eckstein2010}
\begin{align}\label{eq:ppse}
		\tilde \Sigma^\text{NCA}(t,t')=&\, i\sum_\sigma\Big[\tilde c_\sigma^\dagger \tilde G(t,t')\tilde c_\sigma(t')\Delta_\sigma(t,t')\nonumber\\
		&-\tilde c_\sigma(t)\tilde G(t,t')\tilde c_\sigma^\dagger(t')\Delta_\sigma(t',t)\Big],
\end{align}
where the tildes indicate matrix objects in pseudo-particle space. $\tilde\Sigma$ is the pseudo-particle self-energy, $\tilde G$ the pseudo-particle Green's function and $
\tilde c$ is the matrix representation of the impurity annihilation operator in pseudo-particle space. Assuming that the hybridization function decays sufficiently fast, as it is typically the case in insulating systems, one can implement the truncation by  setting the hybridization function to zero for $|t-t'|>t_{\text{cut}}$. Details related to the implementation of the truncation scheme can be found in Ref.~\onlinecite{stahl2022}. 

\section{RESULTS}
\label{sec:results}
\subsection{Equilibrium spectra} 
\begin{figure}
\begin{center}
\includegraphics[width=0.49\textwidth]{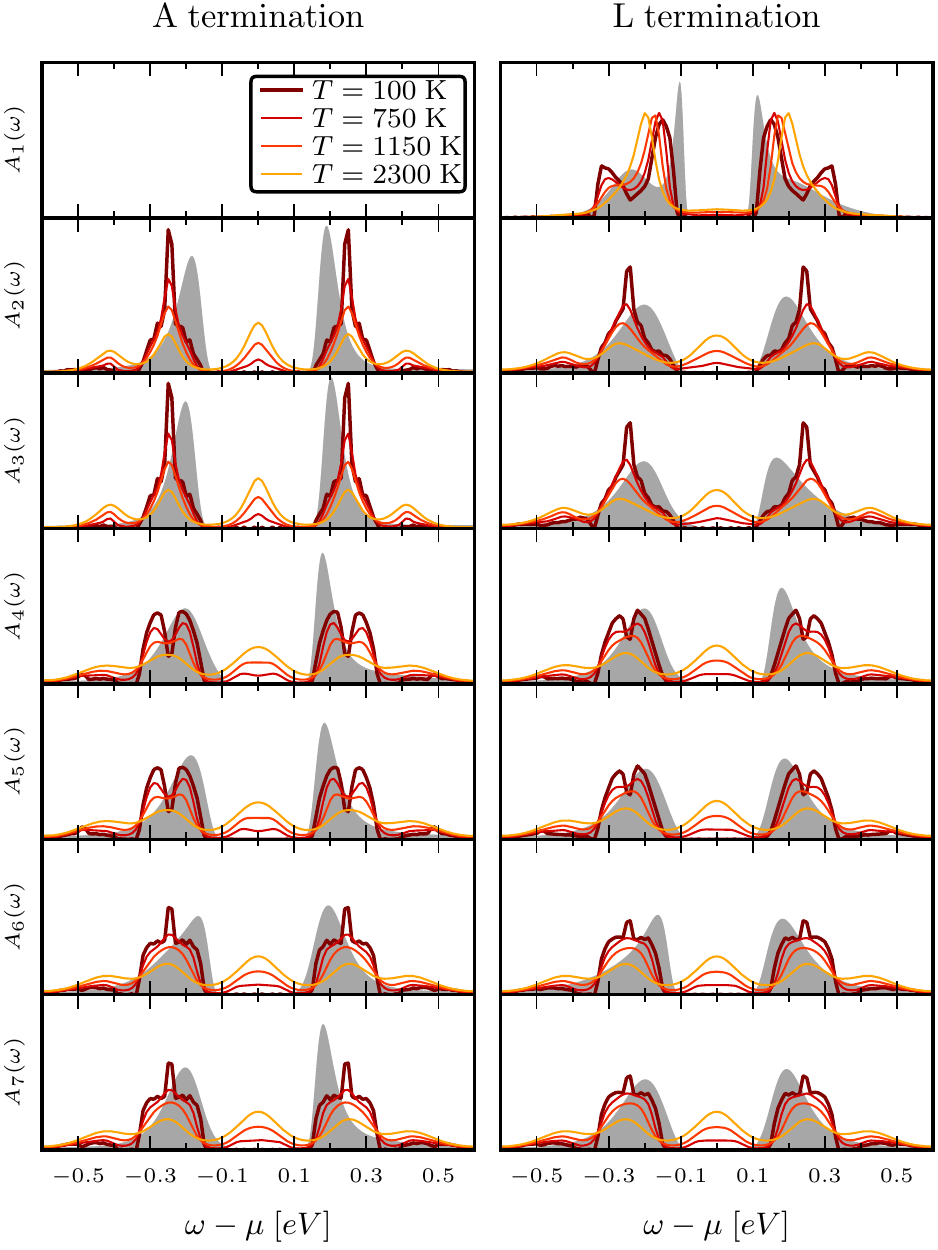}
\caption{ 
Colored lines: Equilibrium layer-resolved local spectral function obtained with the simplified self-consistency at the indicated electronic temperatures. Grey shaded regions: Full $GW$+EDMFT results for the same structures and $T=$ 30 K taken from Ref.~\onlinecite{Petocchi2022}. Note the slightly smaller gap, which is due to screening effects that are not included in the models studied in this work. \label{Figure: Equilib}}
\end{center}
\end{figure}

We simulate a layered setup consisting of eight bilayers and also consider the case of a system terminating with a broken bilayer by adding an additional single-site impurity problem hybridized with the cluster representing the top-most bilayer. The local Hubbard interaction is set to $U=0.4$~eV, since this value was shown to provide an adequate description of the electronic structure in Refs.~\onlinecite{Petocchi2022,Chen2022}. The equilibrium spectral functions $A\left(\omega\right)=-\mathrm{Im} G^R\left(\omega\right)/\pi $, obtained by Fourier transforming the retarded Green's functions of the initial equilibrium system, are reported in Fig.~\ref{Figure: Equilib} for different temperatures $T$.
By comparing the low-temperature spectra for the simplified self-consistency in Fig.~\ref{Figure: Equilib} with those computed using the exact lattice self-consistency and an exact impurity solver\cite{Werner2010} (gray shaded regions),\cite{Petocchi2022} one notices a remarkable agreement, both in terms of the gap sizes and spectral weight distributions. In particular, the characteristic differences between the surface spectra for the two types of terminations are reproduced. From this comparison one can conclude that the model solved with the simplified self-consistency approach captures the balance between kinetic and potential energy and represents a suitable methodology for the following nonequilibrium investigation. 

A striking feature, which even in equilibrium exemplifies the different nature of the insulating states appearing in the structure, is the occurrence of in-gap states exclusively in the bilayers as the temperature is increased. This is a consequence of the different local physics associated with isolated sites as compared to those forming bonding-antibonding states. In the latter case, the addition of an electron (or a hole) to a given site of the dimer creates a state inside the gap, and these states can be thermally populated. In the case of a Mott gap, no such in-gap states appear. The same physics explains the differences in the surface $A\left(\omega\right)$ for the two possible terminations: for the L terminating structure the peaks closer to the Fermi level at $\omega\pm0.15$~eV stem from the interaction-splitted quasiparticle band, while the higher-energy peaks at $\omega\pm0.3$~eV are bonding/antibonding states located at a slightly higher energy with respect to those present in the A terminating system at $\omega\pm0.25$~eV.
\subsection{Energy absorption and diffusion}
\begin{figure}
\includegraphics[width=0.49\textwidth]{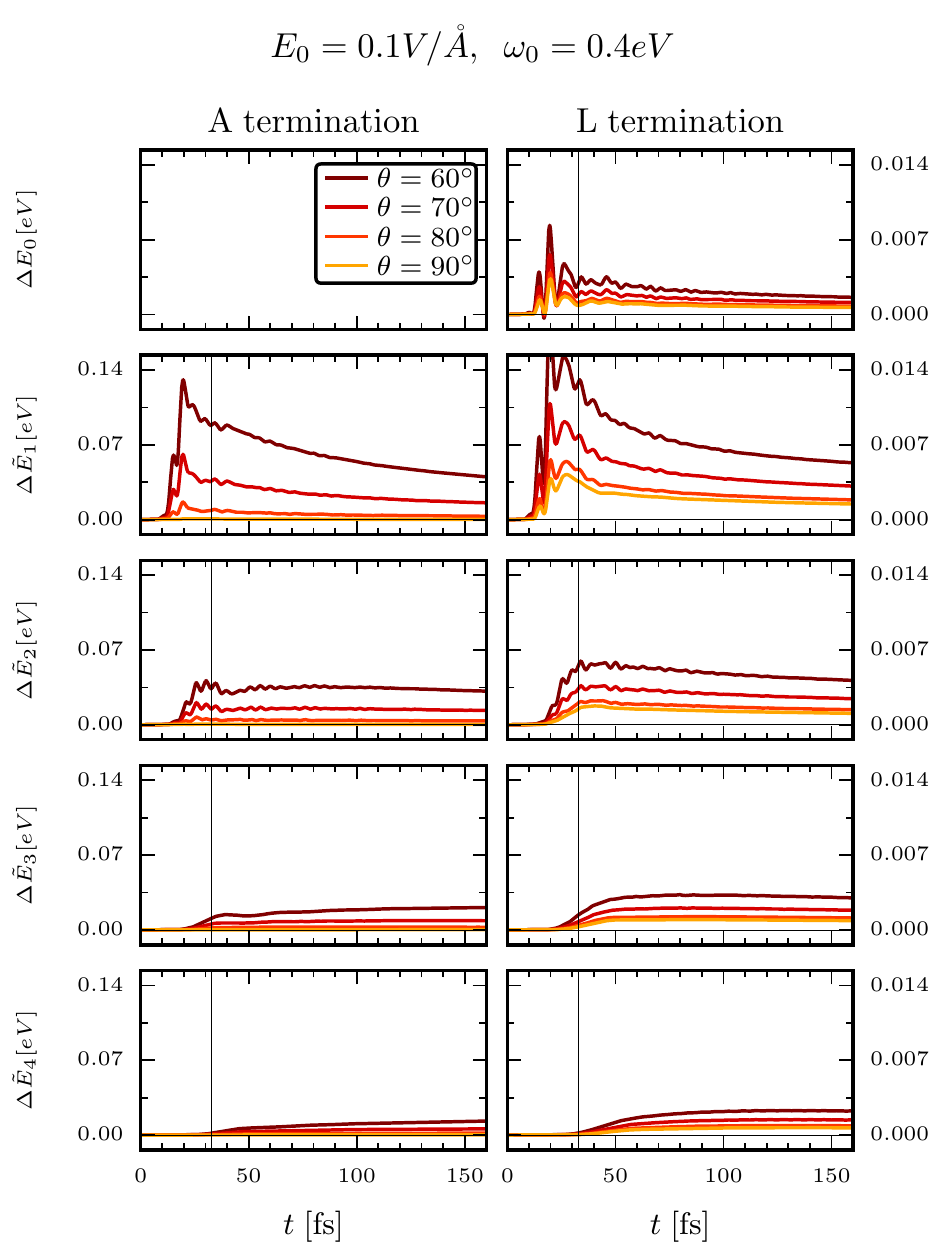}
\caption{ Total energy variation in the different bilayers for the indicated out-of-plane light polarizations. Each panel refers to a given bilayer and the corresponding energy variation is denoted by $\Delta\tilde{E}_i\left(t\right)$, with the exception of the single layer of the L terminating system which corresponds to $\Delta E_0\left(t\right)$. In this figure we report only the region close to the surface of the two 16 layer-thick setups. The end of the pulse is marked in each panel by a vertical black line. \label{Figure: Etot}}
\end{figure}

All the nonequilibrium calculations are performed at an initial inverse temperature $\beta=100\, \mathrm{eV}^{-1}$ ($T=110$ K) using $U=0.4$~eV. Given the position of the peaks of  $A\left(\omega\right)$ we choose the pulse frequencies for the nonequilibrium simulations in the range $0.3 \le \omega_{0} \le 0.9$ eV. The field amplitude is set to $E_0=0.1$ V/\AA \, and we place the center of the pulse at $t_0=16.5$~fs. We simulate a short pulse corresponding to an envelope with whm of $9$~fs. The total energy associated with the different bilayers is given by the sum of the three terms listed in Eq.~\eqref{eqn:energy} and its variation with respect to the equilibrium value $\Delta E_i\left(t\right)$ is reported in Fig.~\ref{Figure: Etot} for different choices of the out-of plane component $\mathbf{p}_z=\mathbf{e}_z\cos\theta $ of the electric field. 
The energy absorption reflects the depth-dependence of our excitation: only the first few layers of the system exhibit an increase in total energy during the pulse, which ends at the time marked by the vertical black line. In the fourth bilayer and below the energy increase is not directly triggered by the pulse, but occurs with a time delay. In particular, the energy increase in these deeper layers occurs in a time window where the total energy in the near-surface layers is already decreasing. This behavior indicates an energy flow from the photo-excited surface region into the bulk. The light polarization of the pulse is found to have a significant effect. When the pump field has a component in the stacking direction $\left(\cos\theta>0\right)$, the energy absorption is significantly increased, especially in the bilayers. We also notice that, for the same pulse amplitude, the energy absorption in the surface Mott layer is substantially smaller and the energy within this layer decays more slowly after the pulse than in the top-most bilayer. It is also noteworthy that the presence of the surface Mott layer significantly reduces the energy absorption in the neighboring bilayer, even if one accounts for the effect of the depth-dependent pump profile.  

\begin{figure}
\includegraphics[width=0.49\textwidth]{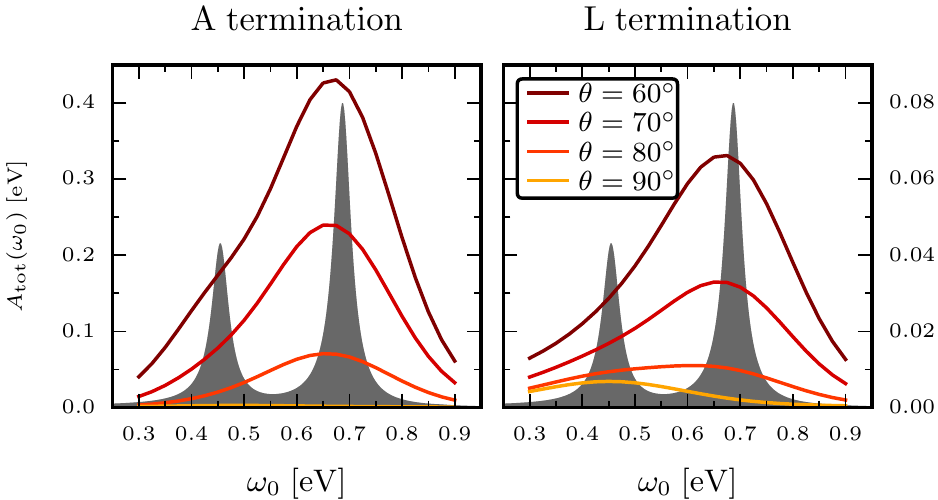}\vspace{5mm}
\includegraphics[trim={-5mm 0 5mm 0},width=0.4\textwidth]{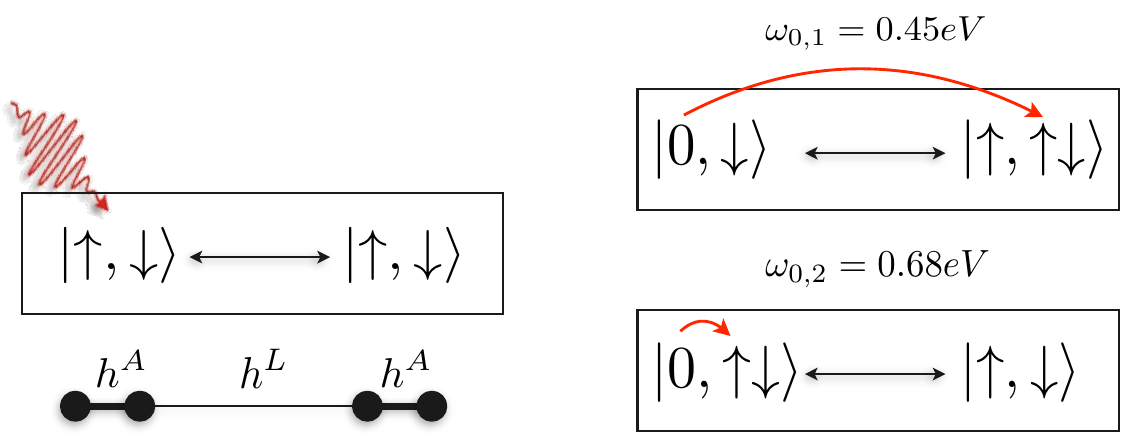}
\caption{ Upper panels: total absorption of the layered setup as a function of the pulse frequency $\omega_0$ and polarization angle $\theta$. The in-plane polarization corresponds to $\theta=90^\circ$ and the shaded region shows the real part of the optical conductivity of the 4-site model. Lower panel: sketch of the minimal 4-site model and the possible hopping processes associated with the two dominant energy excitations.  \label{Figure: Absorption}}
\end{figure}

We provide a comprehensive picture of how the system absorbs energy by performing a scan over the incident light frequency $\omega_0$ and polarization. The total absorption $A_{\mathrm{tot}}\left(\omega_0\right)$ is measured by averaging $\Delta\tilde{E}_i\left(t\right)$ over $\sim10$~fs after the end of the pulse and by summing the contributions of all the bilayers. In the evaluation of $A_{\mathrm{tot}}\left(\omega_0\right)$ we separately counted every kinetic energy term by storing the $\hat{\Delta}_{\Vert}^{i}$ contribution to Eq.~\eqref{eq_delta} for each layer. In Fig.~\ref{Figure: Absorption} we report the results for the two terminations.
The different maximum magnitudes (notice the scales on the $y$-axis) confirm that the surface Mott state significantly reduces the total absorption. Both panels show similar profiles with a two-peak structure located at $\omega_0\sim0.45$~eV and  $\omega_0\sim0.65$~eV that we identified by fitting $A_{\mathrm{tot}}(\omega_0)$. The relative weight of the two peaks depends on the incident light polarization. 

In order to provide an explanation for the observed absorption profiles we solve the atomic Hamiltonian of a 4-site chain with the same interaction and vertical hopping structure as in the extended system. The cluster model consists of two pairs of sites which are dimerized by the strong hopping $h^\text{A}$, and which are connected by the weaker hopping $h^\text{L}$. This structure, sketched in the lower panel of Fig.~\ref{Figure: Absorption}, mimics the realistic simulation in the limit of vanishing in-plane hopping and captures the characteristic excitation energies associated with electrons perturbing the bonding-antibonding states hosted by the bilayers (with overall density fixed to four electrons). The solution of the model at half-filling allows us to compute the real part of the optical conductivity $\mathrm{Re}\sigma_{zz}\left(\omega_{0}\right)$ by means of the exact eigenvalues $\varepsilon_n$ and eigenvectors $\Psi_n$:
\begin{align}
\mathrm{Re}\sigma_{zz}\left(\omega_{0}\right)=&\frac{1}{\omega_0}\sum_{mn}\frac{1}{\left(\omega_{0}+\varepsilon_{m}-\varepsilon_{n}\right)^{2}+\eta^{2}}\times \label{eqn:sigmaED} \\
&\left[\left\langle \Psi_{0}|\Psi_{m}\right\rangle \left\langle \Psi_{m}\right|\hat{j}_{z}\left|\Psi_{n}\right\rangle \left\langle \Psi_{n}\right|\hat{j}_{z}\left|\Psi_{0}\right\rangle -\right. \nonumber \\
&\left.\left\langle \Psi_{0}\right|\hat{j}_{z}\left|\Psi_{m}\right\rangle \left\langle \Psi_{m}\right|\hat{j}_{z}\left|\Psi_{n}\right\rangle \left\langle \Psi_{n}|\Psi_{0}\right\rangle \right] , \nonumber
\end{align}
where $\hat{j}_{z}=i\sum_{i\sigma}h_i\big(\hat{c}_{i\sigma}^{\dagger}\hat{c}_{i+1\sigma}-h.c.\big)$ is the current operator, $\eta$ a small broadening and $h_i$ the hopping parameter which equals either $h^\text{L}$ or $h^\text{A}$ depending on the connected chain sites. The spectra of the 4-site model reveal dominant excitations at the energies $\omega_{0,1}=\varepsilon_{n}-\varepsilon_{m}=0.45$~eV and $\omega_{0,2}=\varepsilon_{n}-\varepsilon_{m}=0.68$~eV. The analysis of the corresponding eigenvectors shows that the one at $\omega_{0,2}$ is related to excitations which move carriers within the bilayers and destroy the bonding state, while the one at $\omega_{0,1}$ is associated with electrons hopping between different bilayers. These results allow us to deduce that in the realistic simulations, the states available for absorption are at $\omega_{0,2}$ when the excitation is out-of plane and at $\omega_{0,1}$ when the pulse excites electrons between the bilayers or within the plane (the total inplane bandwidth is comparable to $h^\text{L}$). This interpretation is corroborated by the absorption profiles, in particular by the yellow curve of Fig.~\ref{Figure: Absorption}, which refers to pure in-plane excitation and does not have the peak at $\omega_{0,2}$, while for fields with an out-of-plane component, the relative weight between the two resonances is concentrated at  $\omega_{0,2}$.

\begin{figure}
\includegraphics[width=0.49\textwidth]{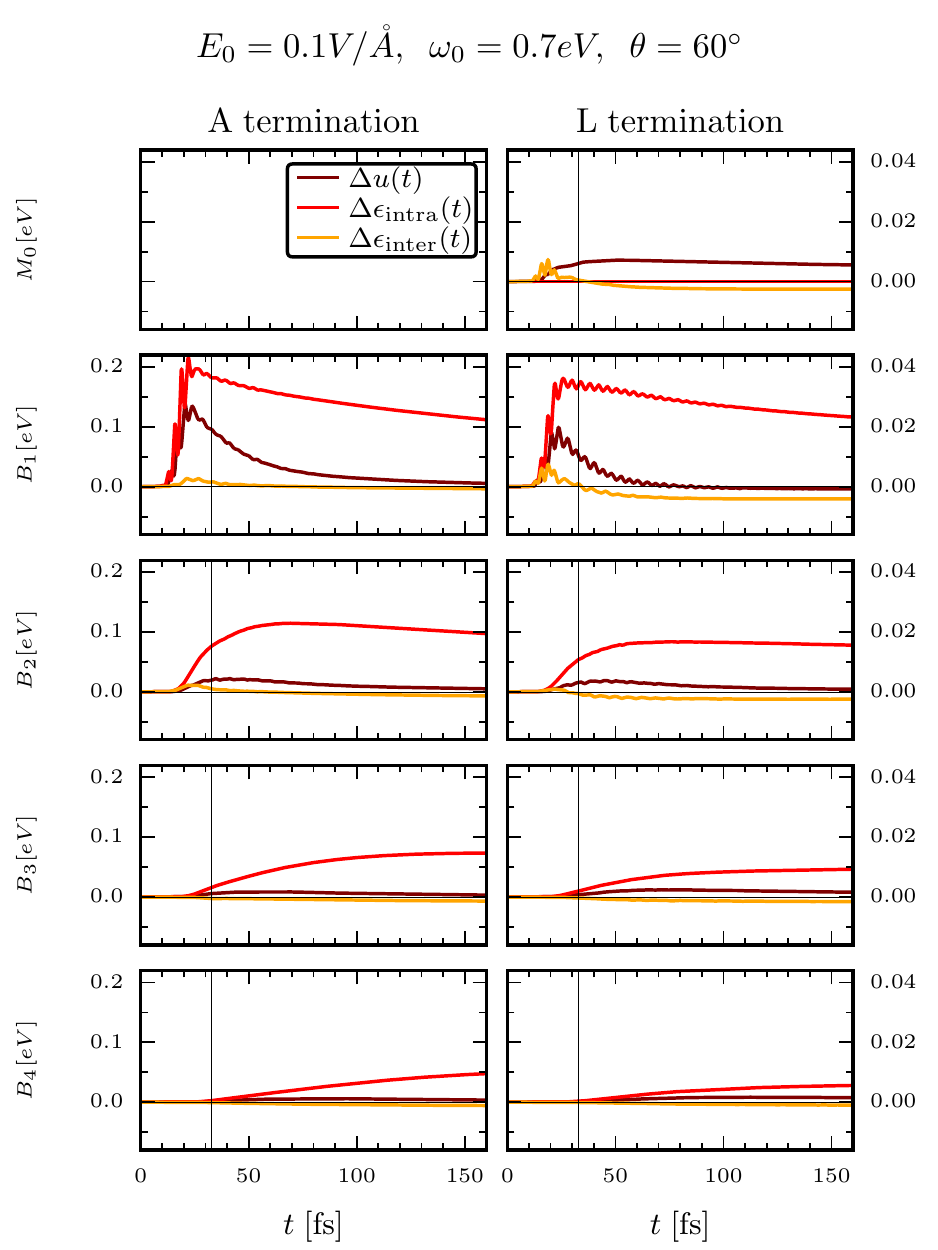}
\caption{ Energy variation subdivided into intra-bilayer potential (brown) and kinetic (red) contributions and inter-bilayer kinetic energy (yellow) for a pulse corresponding to the absorption resonance with highest intensity. \label{Figure: Efrac}} 
\end{figure}

According to Eqs.~\eqref{eqn:energy}-\eqref{eq_ekin_inter} the total energy can be separated into potential and kinetic energy contributions. This is useful since it enables us to identify the nature of the carriers responsible for the energy increase in the system. For the usual case of photo-doped Mott insulators one would expect an {\it increase} in the potential energy as a consequence of the increased doublon density $d\left(t\right)=\left\langle \hat{n}_{\uparrow}\hat{n}_{\downarrow}\right\rangle \left(t\right)$.\cite{eckstein2011} At the same time, an increase in the number of carriers should result in an increased metallicity which manifests itself in a \emph{decrease} of the kinetic energy. In our simulations the kinetic energy is further resolved into contributions coming from the hopping within and between the bilayers. In Fig.~\ref{Figure: Efrac} we report such an analysis for a pulse which corresponds to the strong absorption resonance at $\omega_{0,2}=0.7$~eV.
The first notable observation concerns the potential energy variation $u\left(t\right)$ which is always increased by the pulse but decays very fast in the bilayers with bonding-antibonding gap. Only the surface Mott layer of the L terminating setup exhibits a slow decay, as one would expect for a system with an interaction-to-bandwidth ratio of about one.\cite{eckstein2011} The kinetic energy variation $\epsilon_{\mathrm{inter}} \left(t\right)$ associated with hopping processes within the planes and between the bilayers, indicated by the yellow line in Fig.~\ref{Figure: Efrac}, is always negative at long times, indicating a photo-induced metallic behavior. This behavior is qualitatively the same for all the frequencies and polarizations considered in Fig.~\ref{Figure: Absorption}. Another noteworthy feature of the photo-excited state is the significant positive variation in the kinetic energy internal to the bilayer, $\epsilon_{\mathrm{intra}} \left(t\right)$, which becomes increasingly prominent as $\theta$ decreases. We associate this increase in the (negative) kinetic energy with a \emph{decrease} in the population of the bonding (singlet) state when the system is photo-excited. The effective reduction of the intra-bilayer kinetic energy appears to be mainly responsible for both the strong increase with decreasing $\theta$ in the total energy absorbed by the bilayers, seen in Fig.~\ref{Figure: Etot}, and for the energy diffusion into the bulk. As can be seen from the top layers, especially for the A terminating setup, the doublon recombination occurs within a few tens of femtoseconds after the pulse and the energy diffusion manifests itself mainly in an increase in $\Delta\epsilon_{\mathrm{intra}}$ (disturbance of the intra-bilayer singlets).

\begin{figure}
	\includegraphics[width=0.48\textwidth]{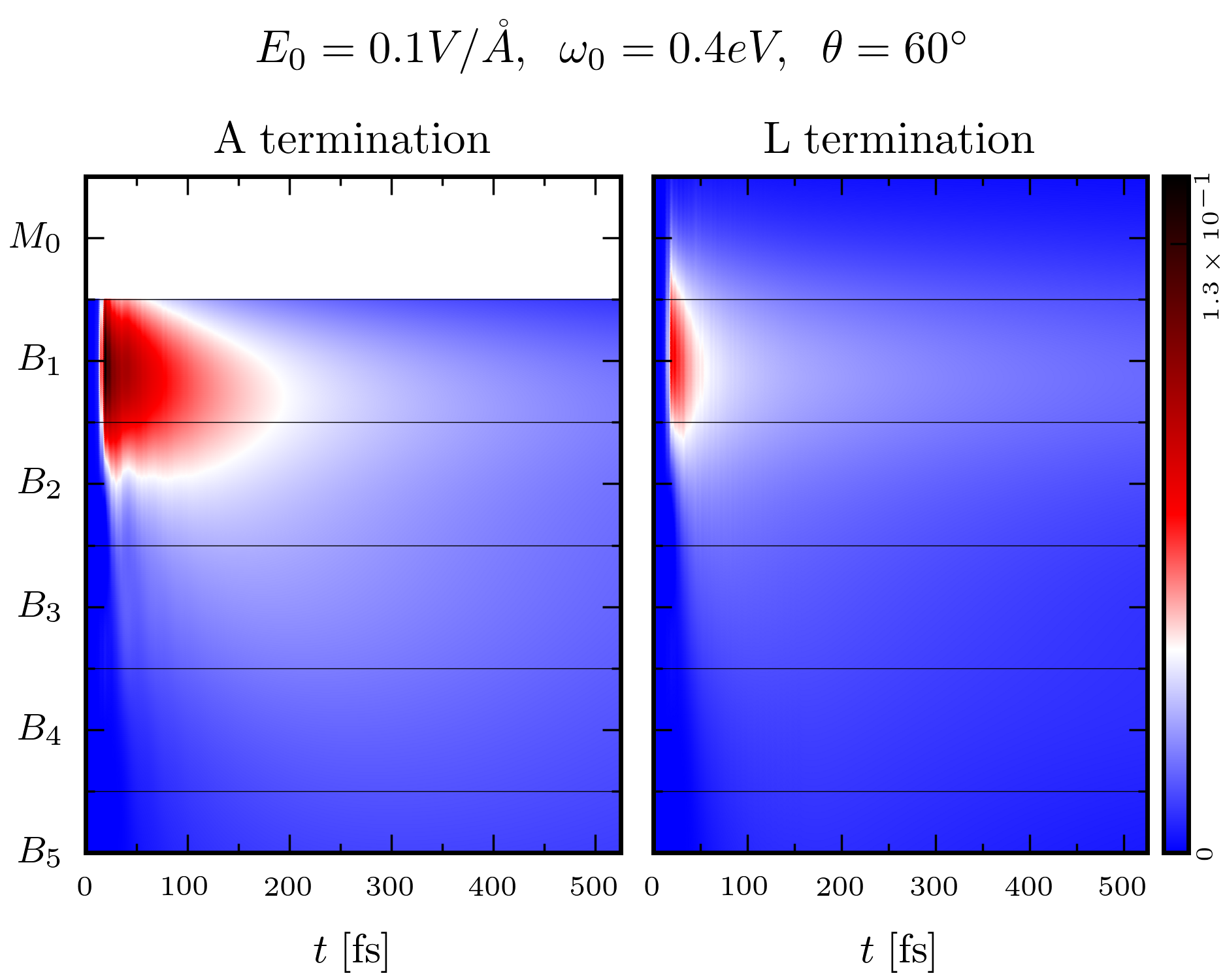}
	\includegraphics[width=0.48\textwidth]{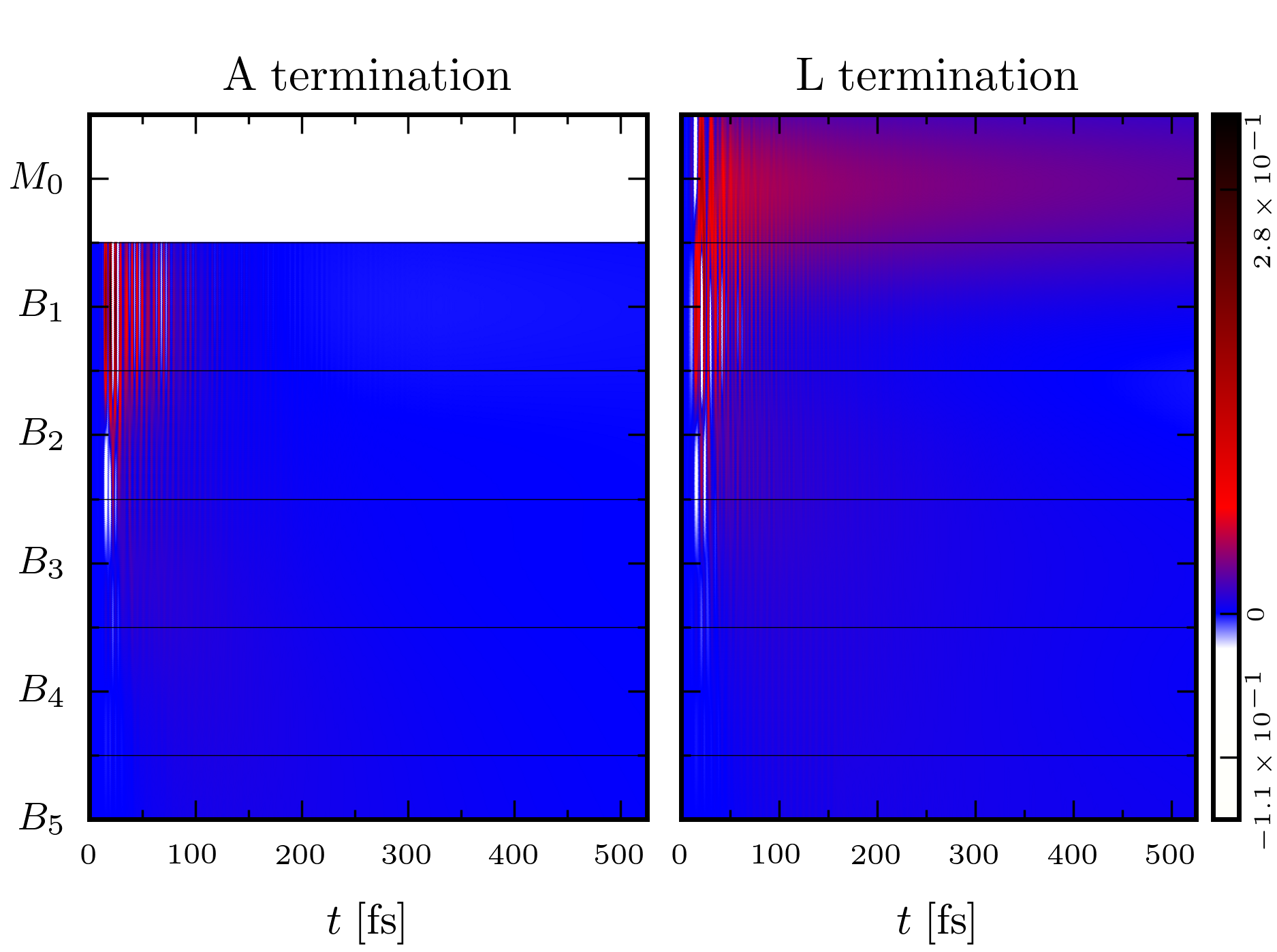}
	\caption{Long-time evolution of the variations in the total energy $E^i=\epsilon_{\mathrm{intra}}^i + u^i + \epsilon_{\mathrm{inter}}^i $ (top) 	
	and doublon density (botton) with respect to the initial equilibrium state for the different bilayers $B_n$ and in the monolayer $M_0$ at the surface of the L terminating setup. In both panels the blue background refers to zero variation. For graphical purposes the variations have been interpolated along the stacking direction and a factor of 4 has been applied to the data for the L termination.\label{fig:diffusion}	}
\end{figure}

In Fig.~\ref{fig:diffusion} we report for both setups the long-time dynamics obtained by applying the memory truncation scheme. The DMFT equations are solved on the full Kadanoff-Baym contour up to $t_{\mathrm{full}}=160$ fs, while at longer times a truncation with $t_\text{cut}=130$ fs is applied. These parameter choices are reasonable since we considered a pulse with whm of 9 fs which results in a smooth dynamics beyond $t_{\mathrm{cut}}$. We confirmed by comparing against the observables computed on the full contour that the results obtained with the truncation scheme are consistent up to five digits, which is the typical precision of our calculations. The agreement can be further improved by increasing the truncation window $t_\text{cut}$. In these simulations, we employ a pulse with  frequency $\omega_{0,1}=0.4$ eV which maximizes the doublon production. In the upper panel of Fig.~\ref{fig:diffusion}  we report the total energy variations $\Delta E_0\left(t\right)$ and $\Delta\tilde{E}_i\left(t\right)$ for each bilayer, similarly to Fig.~\ref{Figure: Etot}. As mentioned above, one notices that the absorption, which corresponds to a positive total energy variation, is largest in the bilayer $B_1$ closest to the surface (note that $B_1$ of the L termination experiences a weaker pulse excitation because of the depth-dependent pump profile) and that this excess energy is transferred to the rest of the structure within 200 fs. The fastest processes appear to be associated with ballistic transport of charge carriers, since the wave fronts have an approximately linear time-dependence. 

The most notable difference between the two terminations is the almost constant small energy increase in bilayer $B_1$ of the L terminating structure. We associate this behavior to the slow diffusion of doublons from the neighboring surface monolayer. To corroborate this interpretation we plot in the lower panels of Fig.~\ref{fig:diffusion} the doublon density variation. For the L termination, we notice a slowly-depleting population of doublons in the surface layer, whereas the doublons detected within the bilayers vanish on the 100 fs timescale. It is also interesting to notice the fact that, in contrast to the total energy, the variation in the doublon density oscillates between positive and negative values (see different limits in the colorbars) when evaluated within a bilayer. This is a consequence of the fact that the double occupation within a layer does not represents a good quantum number in the presence of a strong inter-layer hybidization.
\subsection{Melting of bilayers}
\begin{figure}
\includegraphics[width=0.49\textwidth]{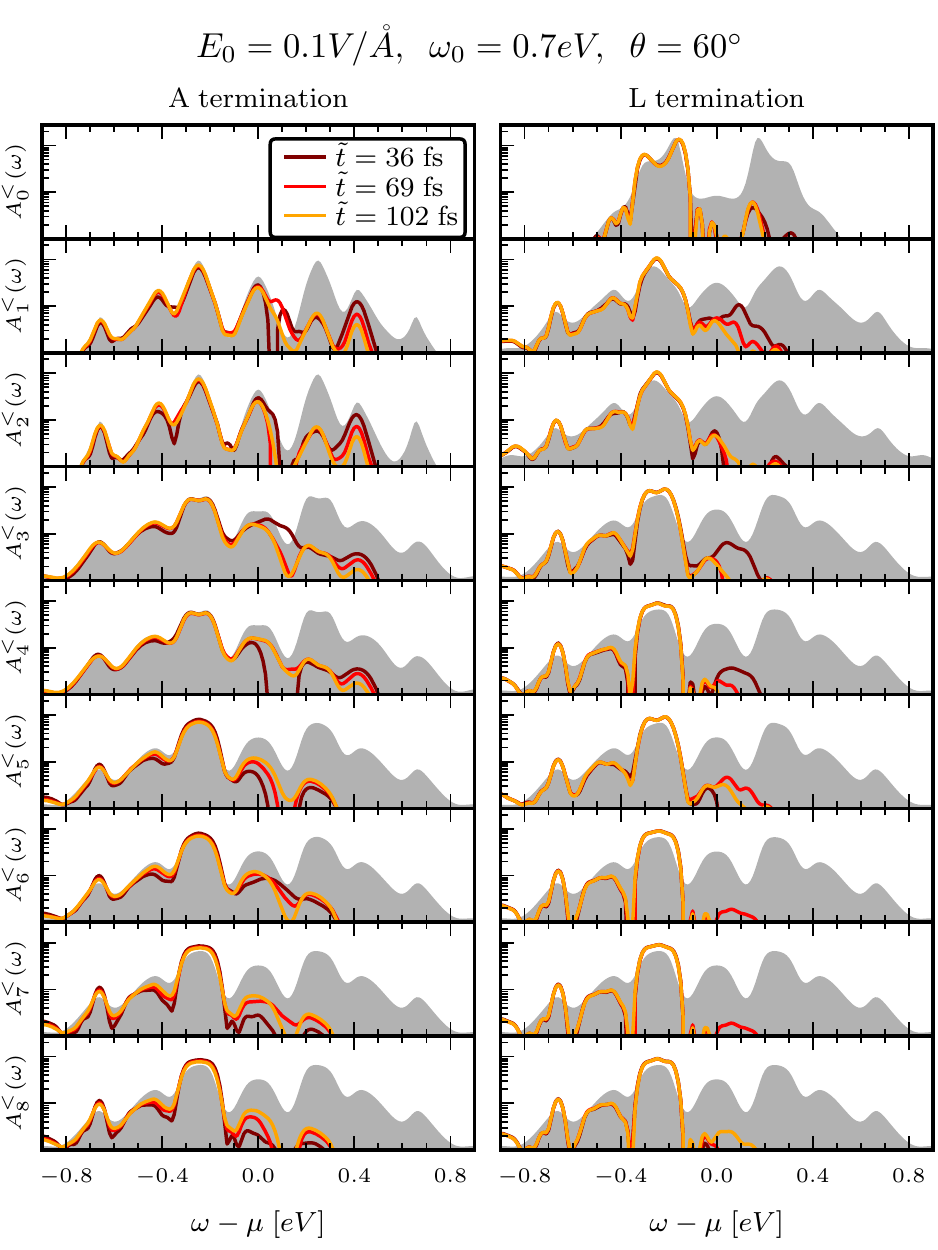}
\caption{ Layer resolved local lesser spectral functions for initial inverse temperature $\beta=100\, \mathrm{eV}^{-1}$ and for the two possible terminations. Different colors refer to different pump-probe delays. For comparison, the grey shaded areas represent the equilibrium spectral functions at $\beta=10\, \mathrm{eV}^{-1}$ computed from the retarded component of the Green's function.  \label{Figure: Snaps}}
\end{figure}

To gain insights into the effects that the suppression of the intra-bilayer bonding state has on the time-dependent occupation we compute the lesser component of the spectral function for each layer,
\begin{equation}
A_{i}^{<}\left(t_{\mathrm{av}}  ,\omega\right)=-\frac{1}{\pi}\mathrm{Im}\int dt_{\mathrm{rel}}e^{i\omega t_{\mathrm{rel}}}G^{i,<}\left(t,t'\right),
\label{Ales}
\end{equation}
where $t_{\mathrm{av}} =\left(t+t'\right)/2$ and $ t_{\mathrm{rel}}=t-t'$. We report the results for different pump-probe delays $\tilde{t}=t_{\mathrm{av}}-t_0$ in Fig.~\ref{Figure: Snaps}, where one notices the emergence of a broad metallic peak which fills the bonding-antibonding gap after the photo-excitation. This band insulator to metal transition is accompanied by a strong heating of the system. The latter  manifests itself through a broadening of the spectra over the whole frequency range, which is highlighted in the figure by plotting the equilibrium spectral function at ten times the temperature ($\beta=10\, \mathrm{eV}^{-1}$) of the initial equilibrium state. 
(The ``nonequilibrium distribution function," obtained by dividing the occupation function by the spectral function, is still very different from a Fermi function, even at $\tilde t=102$~fs, and thus it is not yet possible to define a meaningful temperature, even in the bilayers.) 
We will refer to the presence of these two features, i.e. the broad metallic peak and the increase in the effective temperature, as the \emph{photo-induced melting} of the bilayer. The filling of the gap in the bilayers contrasts with the evolution of the Mott insulating surface layer in the L-terminated system, where one observes an accumulation of doublons at the lower edge of the upper Hubbard band, but no significant change in the gap region. Also, the heating effects seem to be reduced in the Mott layer. 
The trapping of the doublons in the surface layer can be explained by the smaller gap, compared to the bilayers, and the small inter-layer hopping $h^\text{L}=0.045$.

From the study of the Hubbard dimer on the Bethe lattice\cite{Fuhrmann2006} it is known that a reduction of the bonding-antibonding hybridization in the presence of moderate interactions will drive the electronic phase from band insulating to metallic with the emergence of a sharp quasiparticle peak, while Ref.~\onlinecite{Werner2022} discussed the emergence of in-gap states associated with the addition or removal of electrons. We therefore argue that the main effect of pulses with out-of-plane polarization components is twofold: (i) the photo-doped electrons and holes disrupt the bonding state and populate in-gap states, and (ii) the resulting fast electron-hole recombination leads to strong heating. The diffusion that we identified before by the increase in $\epsilon_{\mathrm{intra}}\left(t\right)$ with increasing time and depth of the layer is seen to be associated with this metallic peak. More specifically, if we look at the evolution of the occupation in the deeper layers, one sees that the occupation grows as a function of time in the energy region of the photo-induced metallic peak. In the near-surface spectra, the time evolution instead shows a decrease of the photo-injected charge carriers and their relaxation towards the in-gap states. This is particularly evident for the A terminating setup without the Mott layer, and in the top-most bilayer of the L terminating setup. In agreement with the above discussion of the potential energy variation, the population of the surface Mott layer, however, appears to be almost constant up to the longest simulation times. Some cooling effect is present there, since hot electrons (doublons) can escape to the neighboring layer or be converted into low-energy doublons through impact ionization.\cite{Werner2014,Petocchi2019Hund} 
\subsection{Photoemission}
In experiments the pump pulse is expected to penetrate the sample up to a depth that can be obtained via the Lambert-Beer law and can be estimated to be of the order of several tenths of layers. On the other hand, photoemission spectroscopy (PES) measurements are collecting electrons emitted within their mean-free path, which depends on the kinetic energy they acquire from the photo-excitation. In typical experiments, the signal comes from only the top few layers. In the following, we thus calculate the weighted average

\begin{equation}
O^{\left\{ \mathrm{A},\mathrm{L}\right\} }\left(t,\omega\right)= \sum_{i}A_{i}^{<,\left\{ \mathrm{A},\mathrm{L}\right\} }\left(t,\omega\right)e^{-i/L}, \label{eqn:PES}
\end{equation}
where $A_{i}^{<,\left\{ \mathrm{A},\mathrm{L}\right\} }\left(t,\omega\right)$ is the occupation function \eqref{Ales} of the $i$th layer and $L=0.75$ the mean-free path in units of the inter-layer spacing. 
This quantity is related to the photoemission spectrum, but a quantitative evaluation of the time-resolved PES signal needs to take into account the envelope of the probe pulse\cite{Freericks2009} and the (layer dependent) matrix elements. 
If the duration of the probe pulse is short, this leads to a smearing of the spectral features in time and frequency, compared to the occupied spectral function \eqref{eqn:PES}. Nevertheless, even with an energy and time resolution at the uncertainty limit, the features of interest should remain visible. 
In principle, the occupied spectral function can be extracted from PES measurements with different pulse shapes.\cite{Randi2017} We thus present in the following the results of Eq.~\eqref{eqn:PES} instead of the simulated PES intensity for one fixed pulse shape.

\begin{figure}
\includegraphics[width=0.49\textwidth]{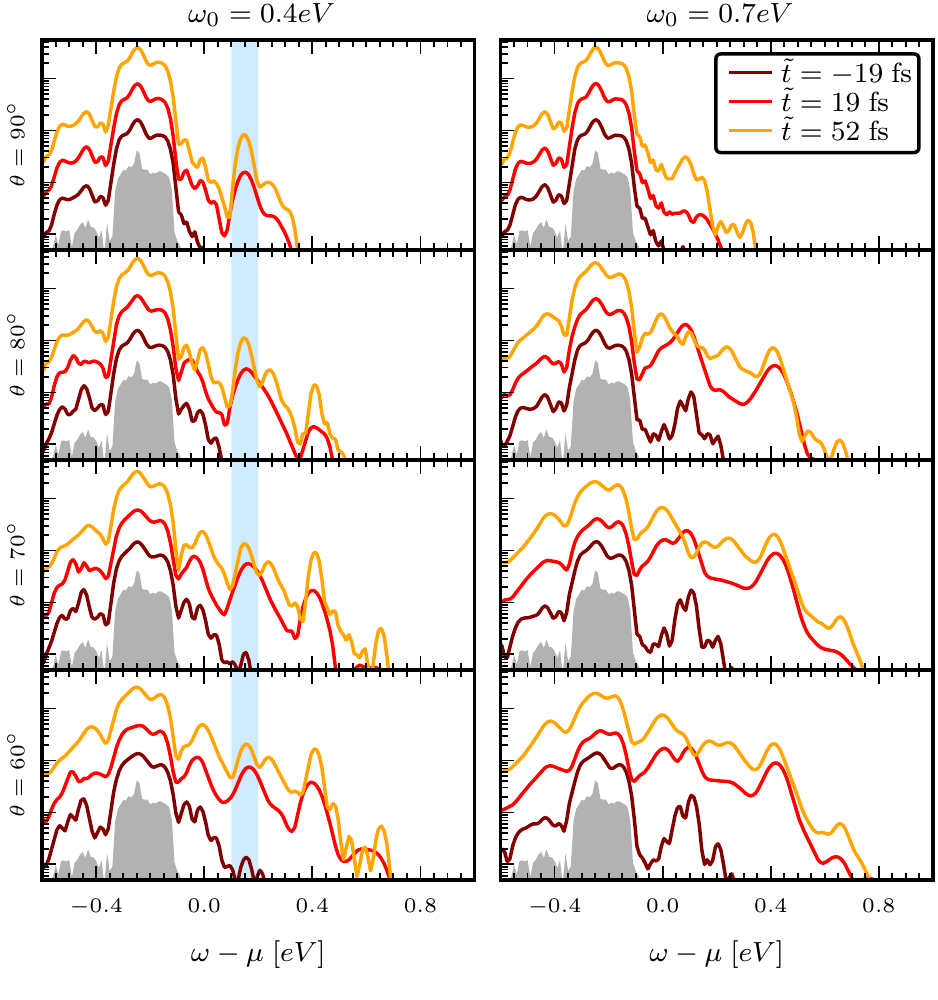}
\caption{
Average of the occupied states close to the sample surface for the A and L terminating systems, for the two absorption resonances and different polarizations. The electric field amplitude is $E_0=0.1$V/\AA \, and the pulse envelope has a whm of 22 fs. Different colors correspond to different pump-probe delays. The gray shaded region shows the occupation of the initial equilibrium state. The curves are offset for better visualization. In light blue we highlight the energy region where the doublon signal is experimentally found. \label{PESv1}}
\end{figure}

\begin{figure}
\includegraphics[width=0.49\textwidth]{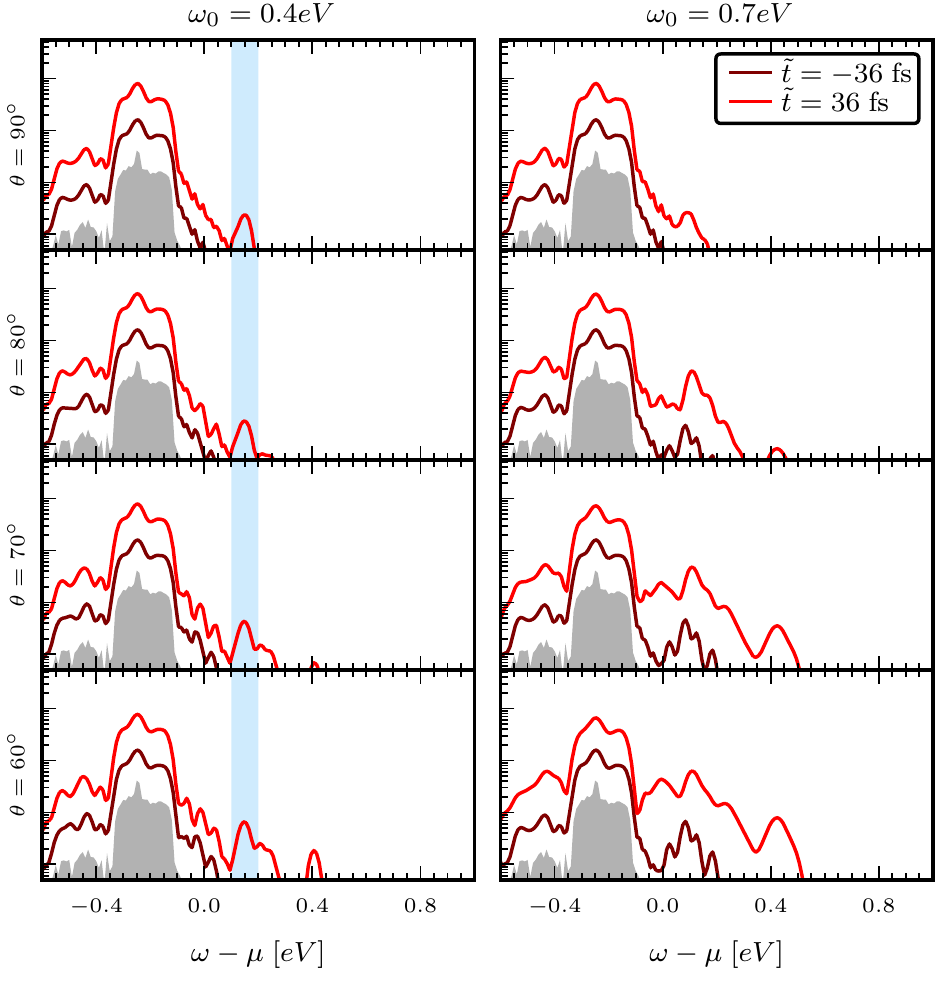}
\caption{Same as Fig.~\ref{PESv1} but with a longer pump pulse with a whm of 55 fs and weaker amplitude $E_0=0.02$ V/\AA. \label{PESv2}}
\end{figure}

It is important to specify the different terminations as the surface spectra, which provide the strongest contribution to $O$, behave differently. Given the large spot size of the probe pulse, it is however reasonable to assume that, in a realistic measurement, the total PES signal originates from domains with different terminations. For this reason we show in Fig.~\ref{PESv1} and \ref{PESv2} the average of the calculated occupied states associated with the two terminations. 

In order to make a closer link to the available pump-probe photoemission experiments performed on the CCDW phase of 1$T$-TaS$_2$, in particular those presented in Ref.~\onlinecite{Ligges2018}, we performed calculations for the same setups described in the previous section but with larger $\sigma$.
We simulated a pulse with whm of 22 fs and amplitude $E_0=0.1$ V/\AA \, in Fig.~\ref{PESv1}, and an even longer pulse with whm of 55 fs (which spans the full non-truncated time window which we can handle) in Fig.~\ref{PESv2}. In this latter case we also decreased the field amplitude to $E_0=0.02$ V/\AA \, in order to be compatible with a fluence of $F\sim0.6\,\mathrm{mJ/cm^2}$. We plot the average of the photoemission signals coming from systems with both terminations after exciting them at the two main resonances $\omega_{0,1}$ and  $\omega_{0,2}$, at different incident angles $\theta$. Negative $\tilde{t}$ refer to snapshots before the pump maximum. 


In all the setups considered, and for all values of $\theta$, we found an emergent peak in the spectra located at $\sim0.2$~eV. This feature is in remarkable agreement with the peak observed in Ref.~\onlinecite{Ligges2018}. If one looks at the curve corresponding to in-plane excitations ($\theta=90^{\circ}$) in Fig.~\ref{PESv1}, it can be noticed that this signal is more pronounced for the lower pump frequency $\omega_{0,1}=0.4$~eV, which is in the optimal range for exciting doublon-hole pairs in the Mott insulating surface state. This feature, which we interpret as the doublon peak of the Mott layer, emerges only for positive $\tilde{t}$.

By decreasing $\theta$ the antibonding states of the bilayers get more and more populated and provide additional contributions to the total PES, which, given the larger gap associated with the bilayers, are located at the higher energy $\sim0.4$~eV. The signal at this energy increases with decreasing $\theta$ and it is stronger for the higher-energy pump-pulse $\omega_{0,2}=0.7$~eV, which confirms that it originates from excitations within the bilayers.

Another noteworthy feature, which is again in close agreement with what has been reported in the experimental paper, is the appearance of a shoulder \emph{before} the pump maximum at $\sim0.1$~eV, which is present only for $\theta<90^{\circ}$ and $\omega_{0,2}$. We associate this signature with excitations of the singlets in the bilayer. This interpretation is confirmed by the layer-resolved time-dependent spectra that we used to compute $O\left(t,\omega\right)$ (not shown).

Finally, we notice that the photo-induced metallic state hosted by the bilayers leads to a continuum of photoemission weight in the whole energy region below 0.4 eV which grows with time and eventually covers the signal at 0.2 eV, at least for small enough $\theta$. This result is compatible with the experimental observation of a prominent ``background" signal, whose intensity increasingly hides the upper Hubbard band as the pump-probe delay increases. In Ref.~\onlinecite{Ligges2018} it was suggested that the observed peak at 175 meV is to be associated with photo-doped doublons which, due to an intrinsic doping of the sample, disappears after 100 fs due to recombination processes. In their analysis Ligges $et~al.$ extracted a 20 fs recombination timescale by separating the doublon signal from the smooth, exponentially decaying background through a fitting procedure. Although this result seems to exclude a long-lived doublon hidden by a growing signal from in-gap states, the fitting procedures may become subtle if the doublon signal becomes very weak. The presence of several non-thermal substructures in the background signal obtained from our model prevents us from performing a reliable estimation of the two timescales by a similar fitting procedure.

Previous theoretical investigations based on nonequilibrium DMFT calculations of the doped single-layer triangular-lattice Hubbard model\cite{Ligges2018} captured the observed timescales for the doublon signature under the assumption of small hole doping, but could not reproduce the gap filling effect.

\section{CONCLUSIONS}
\label{sec:conclusions}
Based on the results presented in this work and in our previous equilibrium study, we propose an alternative interpretation of the photoemission data in Ref.~\onlinecite{Ligges2018}, which explains both the ultrafast disappearance of the signal at 0.2 eV, and the appearance of the in-gap states. Consistent with Ref.~\onlinecite{Ligges2018} we associate the 0.2 eV peak with the upper Hubbard band -- more specifically with the long-lived photo-excited doublons in the Mott insulating surface state of the L terminating system. The spectral weight in the gap region, or background, on the other hand, originates in our multi-layer simulations from the photo-induced disturbance of the singlet states in the bilayers, which make up the A terminating structure and the bulk of the L terminating system. In these bilayers, the photo-excitation creates in-gap states, which leads to a fast recombination of the excited charge carriers, and a strong heating. As the in-gap states get populated and the bilayers heat up, the excitation continuum associated with the bilayers swallows the roughly constant feature coming from the Mott insulating surface state. In this interpretation, we can assume that the system is half-filled and that the doublon population in the upper Hubbard band is long-lived, as one can indeed deduce from Fig.~\ref{Figure: Efrac}. The essential difference to Ref.~\onlinecite{Ligges2018} is that our study builds on the recent insight\cite{Darancet2014,Ritschel2015,Cho2016,Lee2019,Gerasimenko2019,Butler2020,Lee2021,Petocchi2022} that 1$T$-TaS$_2$ is not a pure Mott system, but except for a possibly Mott insulating surface layer, a band insulator whose gap originates from the strong hybridization within bilayers. While other processes beyond our low-energy model may contribute to the background signal, the photo-induced disturbance of the singlet states in the bilayers provides a natural explanation for the appearance of the excitation continuum in time resolved photoemission measurements. 

We also note that a very recent tr-ARPES study\cite{dong2022} on 1$T$-TaS$_2$ provides detailed information on the appearance of the in-gap states during the pump pulse, and confirms the coexistence of a prominent occupation in the in-gap region with a doublon feature around 100 fs after the pump maximum. These results are entirely consistent with the time-resolved spectra reported in our study and the interpretation given here in terms of in-gap states resulting from the melting of the bilayers.

By truncating the time window over which the hybridization is retained using a sufficiently large cutoff, we were able to propagate the interacting solution of our semi-realistic multi-layer model at a computational cost which scales linearly in the maximum time. This allowed us to reach the pico-second timescale and to confirm that the photo-induced excitations in Mott insulating and band-insulating substructures with similar gaps show very different dynamics and timescales. An approximate steady-state with long-lived doublons in the photo-doped surface state and a melted hybridization gap in the top-most bilayer is established after about 200 fs, which corresponds to half an oscillation period of the SOD clusters.\cite{Perfetti2006} On this (and longer) timescales, the coupling of the excited electronic system to the breathing mode in principle needs to be considered. The slow dynamics of the electron-phonon coupled system may be more adequately treated by a different methodological approach, e.~g. based on the quantum Boltzmann equation,\cite{Picano2021} but the present study clarifies the nature of the nonequilibrium state which enters such a scheme as an initial condition. 

It is worth mentioning that the recombination timescale for doublons trapped in the Mott insulating surface layer might be reduced by processes which are not considered in our cluster-DMFT approach, such as the coupling to charge fluctuations. While the multi-layer $GW$+EDMFT analysis of Ref.~\onlinecite{Petocchi2022} includes the feedback from charge fluctuations, this effect is not prominent in the insulating initial state. However, as shown in previous nonequilibrium $GW$+EDMFT studies,\cite{Golez2017,Golez2019} the coupling to charge fluctuations can produce a significant broadening of Hubbard bands in photo-doped states. This could lead to a faster hopping of photo-doped doublons and holons from the surface state into the neighboring bilayer.

The nontrivial interplay between band-insulating and Mott insulating behavior can be systematically explored in time-resolved ARPES by varying the polarization and frequency of the pump pulse. In particular, since the in-gap states result from the disturbance of intra-bilayer singlets, a field polarization in the stacking direction leads to a strong energy absorption and rapid filling of the gap. On the other hand, in-plane polarized pulses can effectively create doublons and holons in the Mott insulating surface layer. Also the pulse frequency can be used to selectively target the different types of excitations, since at least within our model the optimal frequency for the melting of the bilayers is almost a factor of two larger than the optimal frequency for doublon production. Future experiments may exploit these two knobs to shed more light on the photo-carrier dynamics in 1$T$-TaS$_2$. 

A more complete experimental characterization of the photo-doped state could be obtained by combining time-resolved photoemission studies with alternative probes which provide insights into the dynamics of local state populations. Promising approaches to estimate the doublon lifetime and the population of excited dimers are core level X-ray photoemission spectroscopy,\cite{XPS} X-ray absorption spectroscopy,\cite{Baykusheva2022} or resonant inelastic X-ray scattering,\cite{Ament2011} for which theoretical descriptions based on nonequilibrium DMFT have recently been developed.\cite{XAS,Eckstein2021,Werner2021,RIXS2}

\section{ACKNOWLEDGMENTS}
We thank Claude Monney for helpful discussions. F.P. and P.W. acknowledge support from the Swiss National Science Foundation through NCCR MARVEL and from the European Research Council through ERC Consolidator Grant 724103. J. C. is supported by the SNSF through the German research unit QUAST and M.E. also acknowledges funding by the German research unit QUAST. The calculations were performed on the Beo05 clusters at the University of Fribourg.
\bibliography{paper}
\end{document}